\documentclass[letterpaper,english,american,prx,twocolumn,amsmath,amssymb,showpacs,superscriptaddress]{revtex4-1}
\usepackage[T1]{fontenc}
\usepackage{color}
\usepackage{babel}
\usepackage{prettyref}
\usepackage{float}
\usepackage{amsbsy}
\usepackage{amstext}
\usepackage{graphicx}
\usepackage[unicode=true,
 bookmarks=true,bookmarksnumbered=false,bookmarksopen=false,
 breaklinks=false,pdfborder={0 0 1},backref=false,colorlinks=false]
 {hyperref}
\hypersetup{pdftitle={Spectral properties of excitable systems subject to colored noise},
 pdfauthor={Schuecker, Diesmann, Helias},
 pdfsubject={Manuscript for PRL},
 pdfkeywords={colored noise, integrate-and-fire neuron, transfer function},
 colorlinks=true,linkcolor=black,citecolor=black,urlcolor=black,filecolor=black}
\usepackage{breakurl}

\makeatletter

\newrefformat{cap}{\hyperref[#1]{Figure~\ref{#1}}}
\newrefformat{fig}{\hyperref[#1]{Figure~\ref{#1}}}
\newrefformat{tab}{\hyperref[#1]{Table ~\ref{#1}}}
\newrefformat{sec}{\hyperref[#1]{Section~\ref{#1}}}
\newrefformat{sub}{\hyperref[#1]{Section~\ref{#1}}}
\newrefformat{cha}{\hyperref[#1]{Chapter~\ref{#1}}}
\newrefformat{app}{\hyperref[#1]{Appendix~\ref{#1}}}

\makeatother

\begin{document}
\global\long\def\erf{\mathrm{erf}}
\global\long\def\LN{\mathcal{L}_{0}}
\global\long\def\LO{\mathcal{L}_{1}}
\global\long\def\mV{\:\mathrm{mV}}
\global\long\def\ms{\:\mathrm{ms}}
\global\long\def\Hz{\:\mathrm{Hz}}
\global\long\def\Ex{\mathcal{E}}
\global\long\def\In{\mathcal{I}}

\title{Modulated escape from a metastable state driven by colored noise}

\author{Jannis Schuecker }

\affiliation{Institute of Neuroscience and Medicine (INM-6) and Institute for
Advanced Simulation (IAS-6) and JARA BRAIN Institute I, Jülich Research
Centre, Jülich, Germany}

\author{Markus Diesmann }

\affiliation{Institute of Neuroscience and Medicine (INM-6) and Institute for
Advanced Simulation (IAS-6) and JARA BRAIN Institute I, Jülich Research
Centre, Jülich, Germany}

\affiliation{Department of Psychiatry, Psychotherapy and Psychosomatics, Medical
Faculty, RWTH Aachen University, Aachen, Germany}

\affiliation{Department of Physics, Faculty 1, RWTH Aachen University, Aachen,
Germany}

\author{Moritz Helias}

\affiliation{Institute of Neuroscience and Medicine (INM-6) and Institute for
Advanced Simulation (IAS-6) and JARA BRAIN Institute I, Jülich Research
Centre, Jülich, Germany}

\affiliation{Department of Physics, Faculty 1, RWTH Aachen University, Aachen,
Germany}

\date{\today}

\begin{abstract}
Many phenomena in nature are described by excitable systems driven
by colored noise. The temporal correlations in the fluctuations hinder
an analytical treatment. We here present a general method of reduction
to a white-noise system, capturing the color of the noise by effective
and time-dependent boundary conditions. We apply the formalism to
a model of the excitability of neuronal membranes, the leaky integrate-and-fire
neuron model, revealing an analytical expression for the linear response
of the system valid up to moderate frequencies. The closed form analytical
expression enables the characterization of the response properties
of such excitable units and the assessment of oscillations emerging
in networks thereof.
\end{abstract}
\maketitle

\section{Introduction}

In his pioneering work \citet{Kramers1940} investigated chemical
reaction rates by considering the noise-activated escape from a metastable
state as a problem of Brownian motion\textcolor{black}{{} over a barrier.}\textcolor{blue}{{}
}In the overdamped case, this generic setting applies to various phenomena
ranging from ion channel gating \citep{Goychuk02_3552}, Josephson
junctions \citep{malakhov96_64,pankratov04_177001,mantegna00_3025},
tunnel diodes \citep{Mantegna96_563}, semiconductor lasers \citep{hales00_87}
to abstract models of cancer growth \citep{spagnolo08}. These studies
assume the noise statistics to have a white spectrum. This idealization
simplifies the analytical treatment and is the limit of non-white
processes for vanishing correlation time \citep{Gardiner83}. White
noise cannot exist in real systems, which is obvious considering,
for example, the voltage fluctuations generated by thermal agitation
in a resistor \citep{Johnson28,Nyquist28}: if fluctuations had a
flat spectrum, power dissipation would be infinite. In a non-equilibrium
setting a system is driven by an external fluctuating force. The simplest
model for such a non-white noise is characterized by a single time
constant, representing a more realistic colored-noise model. 

The effect of colored noise is relevant in modeling the phase difference
in lasers \citep{Vogel95}, the velocity in turbulence models \citep{Frisch81_2673},
genetic selection \citep{Horsthemke84} and chemical reactions \citep{Lindenberg83}
(for further examples see \citep{Moss89_2,Moss89_3}). Its influence
on the escape from a metastable state has been studied theoretically
as the stationary mean first passage time (MFPT) of a particle in
the Landau potential \citep{Sancho82,Lindenberg83,Haenggi85,Fox86,Grigolini86,Doering87}.
Treating colored noise analytically comes along with considerable
difficulties, since it adds a dimension to the governing Fokker-Planck
equation (FPE) and the common strategy is to reduce the colored-noise
problem to an effective white-noise system. \citet{Doering87} and
\citet{Klosek98} developed these approaches further by singular perturbation
methods and boundary-layer theory, showing that the leading order
correction to the static MFPT stems from an appropriate treatment
of the boundary conditions of the effective system.

The organization of this work is as follows. In \prettyref{sec:Reduction-from-colored}\textcolor{black}{{}
we extend the works by }\citet{Doering87} and \citet{Klosek98}\textcolor{black}{{}
to the time-dependent case by a perturbation expansion of the flux
operator appearing in the FPE itself. }This approach leads to a general
method that reduces a first order differential equation driven by
additive and fast colored noise to an effective one-dimensional system,
allowing the study of time-dependent phenomena and thus revealing
the spectral properties of the system. The effective formulation implicitly
contains the matching between outer and boundary-layer solutions.
The latter appear close to absorbing boundaries and are obtained by
a half-range expansion. Our main result is that colored-noise approximations
for stationary but more importantly also for dynamic quantities are
directly obtained by shifting the location of the boundary conditions
in the solutions for the corresponding white-noise system.

In \prettyref{sec:LIF-neuron} we apply this general result to the
particular problem of modeling biological membranes: the leaky integrate-and-fire
(LIF) neuron model \citep{Lapicque07,Stein67a} with exponentially
decaying post-synaptic currents can equivalently be described as a
first-order differential equation driven by colored noise. Based on
the works of \citet{Doering87} and \citet{Klosek98}, \citeauthor{Brunel01_2186}
\citep{Brunel01_2186,Fourcaud02} calculated the high-frequency limit
of its transfer function. Here we complement these works deriving
a novel analytical expression for the transfer function valid up to
moderate frequencies, which we confirm by direct simulations. While\textcolor{black}{{}
for slow noise, an adiabatic approximation for the transfer function
is known \citep{Morenobote06_028101},} this first order correction
in the time scale of the noise is a qualitatively new result. In the
previous work by \citet{Fourcaud02}\textcolor{blue}{{} }\textcolor{black}{it
was shown that the first order correction vanishes in case of the
integrate-and-fire neuron without a leak term, i.e. for the perfect
integrator (PIF). For this type of neuron, the marginal statistics
of the spike train was characterized for an arbitrary time scale of
the noise with weak amplitude \citep{Lindner04_0229011}. Further
extensions to multiple time scales have recently been developed \citep{schwalger15_1}.
However, to date the correction to the transfer function of the LIF
model is unknown in the }biologically relevant regime of moderate
frequencies. 

The transfer function is at the heart of the contemporary theory of
fluctuations in spiking neuronal networks \citep{Lindner05_061919,Shea-Brown08}
and allows the derivation of analytical expressions for experimentally
observable measures, such as pairwise correlations \citep{Trousdale12_e1002408}
and oscillations in the population activity \citep{Brunel99,Brunel03a}.
In \prettyref{sec:Balanced-Random-Network} we show analytically how
realistic synaptic filtering affects the power spectrum in a recurrent
network of excitatory and inhibitory neurons, a question that has
been inaccessible prior to the present work.

\section{Reduction from colored to white noise\label{sec:Reduction-from-colored}}

\begin{figure}
\centering{}\includegraphics{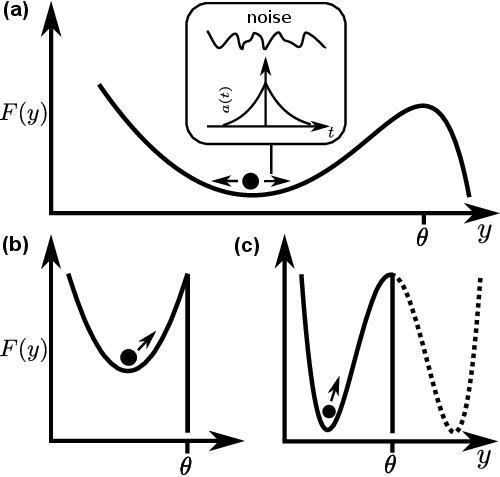}\caption{\label{fig:potential}\textbf{Escape from a metastable state}. \textbf{A
}Overdamped motion of a particle driven by colored noise (with exponentially
decaying autocovariance $a(t),$ inset) in a quadratic potential $F(y)=y^{2}$
with smooth boundary near $\theta$. \textbf{B }Quadratic potential
with hard absorbing boundary. \textbf{C }Bistable potential $F(y)=-y^{2}+y^{4}$
with absorbing boundary at the maximum.}
\end{figure}

Consider a pair of coupled stochastic differential equations (SDE)
with a slow component $y$ with time scale $\tau$, driven by a fast
Ornstein-Uhlenbeck process $z$ with time scale $\tau_{s}$. In dimensionless
time $s=t/\tau$ and with $k=\sqrt{\tau_{s}/\tau}$ relating the two
time constants we have

\begin{eqnarray}
\frac{dy}{ds} & = & f(y,s)+\frac{z}{k}\nonumber \\
k\frac{dz}{ds} & = & -\frac{z}{k}+\xi,\label{eq:diffeq_general}
\end{eqnarray}
with a unit variance white noise $\langle\xi(s+u)\,\xi(s)\rangle=\delta(u)$.
The setting \prettyref{eq:diffeq_general} describes an overdamped
Brownian motion of a particle in a possibly time dependent potential
$F(y,s)=-\int^{y}f(y^{\prime},s)\,dy^{\prime}$, driven by colored
noise. The generic problem of escape from such a potential is illustrated
in \prettyref{fig:potential}A where the particle has to overcome
a smooth barrier. The latter can be simplified, assuming an absorbing
boundary at the right end $y=\theta$ of the domain, illustrated in
\prettyref{fig:potential}B for the quadratic potential $F(y)=y^{2}$.
 The potential $F(y)=-y^{2}+y^{4}$ (\prettyref{fig:potential}C)
is regarded as the archetypal form giving rise to a bistable system
and is reviewed in \citep{haenggi90_251} and \citep{haenggi95_239}.
The MFTP from one well to the peak of the potential can be calculated
by assuming an absorbing boundary at the maximum, formally reducing
this case to the one shown in \prettyref{fig:potential}B. Note that
the MFTP from one well to the other is not directly related to the
MFTP to the peak, but rather is twice the MFTP to an absorbing boundary
on the separatrix in the two-dimensional domain \citep{Haenggi88}
spanned by $y$ and $z$. However, we are here interested in the case
where the component $y$ crosses a constant threshold value \citep{Doering88}.

Initially we revisit the intuitive argument that in the case of fast
noise $\tau_{s}\ll\tau$ the slow component $y$ approximately obeys
the one-dimensional SDE $\frac{dy}{ds}=f(y,s)+\xi(s)$. Heuristically
this can be viewed as $y$ integrating $z$ on a time scale $\tau$.
For finite $\tau_{s}$ the integral of the autocorrelation of $z$
is $k^{2}$, identical to the limit of vanishing $\tau_{s}$, where
$z(s)=k\xi(s)$ becomes a white noise. We here set out to formally
derive an effective diffusion equation for $y$ to obtain a formulation
in which we can include the absorbing boundaries that are essential
to study an escape problem. We consider the FPE \citep{Risken96}
corresponding to the two-dimensional system \prettyref{eq:diffeq_general}
\begin{equation}
k^{2}\partial_{s}P=\partial_{z}\left(\frac{1}{2}\partial_{z}+z\right)\,P-k^{2}\partial_{y}S_{y}\,P,\label{eq:FP_2D}
\end{equation}
where $P(y,z,s)$ denotes the probability density and we introduce
the probability flux operator in $y$-direction as $S_{y}=f(y,s)+z/k$.
Factoring-off the stationary solution of the fast part of the Fokker-Planck
operator, $P=Q\,\frac{e^{-z^{2}}}{\sqrt{\pi}}$, we observe the change
of the differential operator \foreignlanguage{english}{$\partial_{z}\left(\frac{1}{2}\partial_{z}+z\right)\to L\equiv\left(\frac{1}{2}\partial_{z}-z\right)\partial_{z}$},
which transforms \prettyref{eq:FP_2D} to

\begin{equation}
k^{2}\partial_{s}Q=LQ-kz\partial_{y}Q-k^{2}\partial_{y}\,f(y,s)\,Q.\label{eq:FP_Q_general}
\end{equation}
We refer to $Q$ as the outer solution, since initially we do not
consider the absorbing boundary condition. The strategy is as follows:
we show that the terms of first and second order in the small parameter
$k$ of the perturbation ansatz 
\begin{equation}
Q=\sum_{n=0}^{2}k^{n}\,Q^{(n)}+O(k^{3})\label{eq:perturbation_ansatz}
\end{equation}
 cause an effective flux acting on the $z$-marginalized solution
$\tilde{P}(y,s)=\int dz\,\frac{e^{-z^{2}}}{\sqrt{\pi}}Q(y,z,s)$ that
can be expressed as a one-dimensional FPE which is correct up to linear
order in $k$. To this end we need to know the first order correction
to the marginalized probability flux $\nu_{y}(y,s)\equiv\int dz\,\frac{e^{-z^{2}}}{\sqrt{\pi}}S_{y}Q(y,z,s)=\sum_{n=0}^{1}k^{n}\,\nu_{y}^{(n)}(y,s)+O(k^{2})$.
Inserting the perturbation ansatz \prettyref{eq:perturbation_ansatz}
into \prettyref{eq:FP_Q_general} we have $LQ^{(0)}=0$. Noting the
property $Lz^{n}=\frac{1}{2}n(n-1)z^{n-2}-nz^{n}$, we see that the
lowest order does not imply any further constraints on the $z$-independent
solution $Q^{(0)}(y,s)$, which must be consistent with the solution
to the one-dimensional Fokker-Planck equation corresponding to the
limit $k\to0$ of \prettyref{eq:diffeq_general}. The first and second
orders are 
\begin{eqnarray}
LQ^{(1)} & = & z\partial_{y}Q^{(0)}\label{eq:perturb_expansion_general}\\
LQ^{(2)} & = & \partial_{s}Q^{(0)}+z\partial_{y}Q^{(1)}+\partial_{y}f(y,s)Q^{(0)}.\nonumber 
\end{eqnarray}
With $Lz=-z$, the general solution for the first order 
\begin{equation}
Q^{(1)}(y,z,s)=Q_{0}^{(1)}(y,s)-z\partial_{y}Q^{(0)}(y,s)\label{eq:Q1_general}
\end{equation}
leaves the freedom to choose a homogeneous ($z$-independent) solution
$Q_{0}^{(1)}(y,s)$ of $L$. To generate the term linear in $z$ on
the right hand side of the second order in \prettyref{eq:perturb_expansion_general},
we need a term $-z\partial_{y}Q^{(1)}$. The terms constant in $z$
require contributions proportional to $z^{2}$, because $Lz^{2}=-2z^{2}+1$.
However, they can be dropped right away since their contribution to
$\,\nu_{y}^{(1)}$ vanishes after marginalization. For the same reason
the homogeneous solution $Q_{0}^{(2)}(y,s)$ can be dropped. Collecting
all terms which contribute to $\nu_{y}$ in orders $k^{2}$ and higher
in $\iota$, the only relevant part of the second order solution is
$-z\partial_{y}Q_{0}^{(1)}(y,s)$, which leaves us with
\begin{equation}
\begin{aligned}Q(y,z,s) & =Q^{(0)}(y,s)+kQ_{0}^{(1)}(y,s)\\
 & -kz\partial_{y}Q^{(0)}(y,s)-k^{2}z\partial_{y}Q_{0}^{(1)}(y,s)+\iota,
\end{aligned}
\label{eq:Q_eff}
\end{equation}
from which we obtain the marginalized flux
\begin{equation}
\nu_{y}(y,s)=\left(f(y,s)-\frac{1}{2}\partial_{y}\right)\tilde{P}(y,s)+O(k^{2}).\label{eq:effective_flux}
\end{equation}
We observe that $f(y,s)-\frac{1}{2}\partial_{y}$ is the flux operator
of a one-dimensional system driven by a unit variance white noise
and $\tilde{P}(y,s)\equiv Q^{(0)}(y,s)+kQ_{0}^{(1)}(y,s)$ is the
marginalization of \prettyref{eq:Q_eff} over $z$. Note that in \prettyref{eq:Q_eff}
the higher order terms in $k$ appear due to the operator $kz\partial_{y}$
in \eqref{eq:FP_Q_general} that couples the $z$ and $y$ coordinate.
Eq. \eqref{eq:effective_flux} shows that these terms cause an effective
flux that only depends on the $z$-marginalized solution $\tilde{P}(y,s)$.
This allows us to obtain the time evolution by applying the continuity
equation to the effective flux \eqref{eq:effective_flux} yielding
the effective FPE

\begin{equation}
\partial_{s}\tilde{P}=-\partial_{y}\nu_{y}(y,s)=\partial_{y}\left(-f(y,s)+\frac{1}{2}\partial_{y}\right)\,\tilde{P}.\label{eq:FP_tilde}
\end{equation}

\begin{figure}
\centering{}\includegraphics{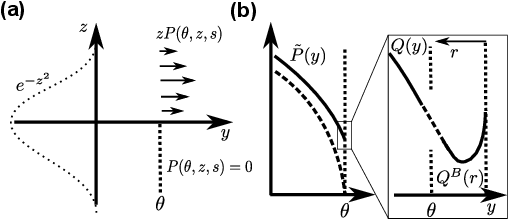}\caption{\label{fig:Illustration-of-boundary}\textbf{Boundary conditions.}
\textbf{A} Flux and boundary condition in the two-dimensional colored-noise
system. On the negative half plane $z<0$ the density must vanish
at threshold $\theta$. Dashed curve: density marginalized over $y$.
\textbf{B} Density of the white-noise system (dashed curve) vanishes
at threshold. The density of the effective system (solid curve) has
a finite value at threshold which is determined by the matching between
the outer solution $Q$ and the boundary layer solution $Q^{B}$ (shown
in enlargement). }
\end{figure}
Effective FPEs, to first order identical to \eqref{eq:FP_tilde},
but also including higher order terms, have been derived earlier
\citep{Sancho82,Lindenberg83,Haenggi85,Fox86,Grigolini86}. These
approaches have been criticized for an untended treatment of the boundary
conditions of the effective system \citep{Klosek98}. \citet{Doering87}
and \citet{Klosek98} used singular perturbation methods and boundary
layer theory to show that the $O(k)$ correction to the static MFPT
stems from colored-noise boundary conditions for the marginalized
density of the effective system. Extending this approach to the transient
case requires a time-dependent boundary condition for $\tilde{P}$
or, equivalently, for $Q_{0}^{(1)}$, because we assume the one-dimensional
problem to be exactly solvable and hence the boundary value of $Q^{(0)}$
to be known. Without loss of generality, we assume an absorbing boundary
at the right end $y=\theta$ of the domain. Thus trajectories must
not enter the domain from above threshold, implying the flux to vanish
at $\theta$ for all points with negative velocity in $y$ given by
$f(\theta,s)+\frac{z}{k}<0$. A change of coordinates $z+kf(\theta,s)\to z$
simplifies the condition to $\frac{z}{k}\,Q(\theta,z,s)=0$ for $z<0$.
The flux and this half range boundary condition are shown in \prettyref{fig:Illustration-of-boundary}A.
The resulting boundary layer at $\theta$ requires the transformation
of the FPE \eqref{eq:FP_Q_general} to the shifted and scaled coordinate
$r=\frac{y-\theta}{k}$, which yields
\begin{equation}
\begin{aligned}k^{2}\partial_{s}Q^{B} & =LQ^{B}-z\partial_{r}Q^{B}\\
 & +k\,G(\theta,r,s,z)\,Q^{B}+O(k^{3}),
\end{aligned}
\label{eq:FP_boundary_variable}
\end{equation}
with $Q^{B}(r,z,s)\equiv Q(y(r),z,s)$ and the operator $G(\theta,r,s,z)=f(\theta,s)\partial_{z}-\partial_{r}\,(f(kr+\theta,s)-f(\theta,s))$.
The boundary condition then takes the form 
\begin{equation}
Q^{B}(0,z,s)=0\quad\forall z<0.\label{eq:BC_boundary_layer}
\end{equation}
With the perturbation ansatz $Q^{B}=\sum_{n=0}^{1}k^{n}\,Q^{B(n)}+O(k^{2})$
we obtain
\begin{eqnarray}
LQ^{B(0)}-z\partial_{r}Q^{B(0)} & = & 0\nonumber \\
LQ^{B(1)}-z\partial_{r}Q^{B(1)} & = & G(\theta,r,s,z)\,Q^{B(0)},\label{eq:order1_boundary}
\end{eqnarray}
the solution of which must match the outer solution. The latter varies
only weakly on the scale of $r$ and therefore a first order Taylor
expansion at the boundary yields the matching condition $Q^{B}(r,z,s)=Q(\theta,z,s)+kr\,\partial_{y}Q(\theta,z,s)$,
illustrated in \prettyref{fig:Illustration-of-boundary}B. We note
with $Q^{(0)}(\theta,s)=0$ that the zeroth order $Q^{B(0)}$ vanishes.
Inserting \eqref{eq:Q1_general} into the matching condition, with
$-\frac{1}{2}\partial_{y}Q^{(0)}(\theta,s)=\nu_{y}^{(0)}(\theta,s)\equiv\nu_{y}^{(0)}(s)$
the instantaneous flux in the white-noise system, the first order
takes the form
\begin{equation}
Q^{B(1)}(r,z,s)=Q_{0}^{(1)}(\theta,s)+2\nu_{y}^{(0)}(s)\,(z-r).\label{eq:QT_boundary_z}
\end{equation}
The vanishing zeroth order implies with \prettyref{eq:order1_boundary}
for the first order $LQ^{B(1)}-z\partial_{r}Q^{B(1)}=0.$ Appendix
B of \citet{Klosek98} states the solution of the latter equation
satisfying the half-range boundary condition
\begin{equation}
Q^{B(1)}(r,z,s)=C(s)\Big(\frac{\alpha}{2}+z-r+\sum_{n=1}^{\infty}b_{n}(z)\,e^{\sqrt{2n}r}\Big),\label{eq:bl_solution}
\end{equation}
with $\alpha=\sqrt{2}|\zeta(\frac{1}{2})|$ given by Riemann's $\zeta$-function
and $b_{n}$ proportional to the $n$-th Hermite polynomial. We equate
\prettyref{eq:bl_solution} to \prettyref{eq:QT_boundary_z} neglecting
the latter exponential term which decays on a small length scale,
so that the term proportional to $z-r$ fixes the time dependent function
$C(s)=2\nu_{y}^{(0)}(s)$ and hence the boundary value $Q_{0}^{(1)}(\theta,s)=\alpha\nu_{y}^{(0)}(s)$.
This yields the central result of our theory: We have reduced the
colored-noise problem \prettyref{eq:diffeq_general} to the solution
of a one-dimensional FPE \prettyref{eq:FP_tilde} with the time-dependent
boundary condition 
\begin{equation}
\tilde{P}(\theta,s)=k\alpha\nu_{y}^{(0)}(s),\label{eq:Boundary_conditions_static}
\end{equation}
where the time-dependent flux $\nu_{y}^{(0)}(s)$ is obtained from
the solution of the corresponding white-noise problem. The boundary
condition can be understood intuitively: the filtered noise slows
down the diffusion at the absorbing boundary with increasing $k$;
the noise spectrum and the velocity of $y$ are bounded, so in contrast
to the white-noise case, there can be a finite density at the boundary.
Its magnitude results from a momentary equilibrium of the escape crossing
the boundary and the flow towards it, where the latter is approximated
by the flow $\nu_{y}^{(0)}$ at threshold in the white noise system.
In principle one can obtain colored-noise solutions combining the
effective system \textcolor{black}{\eqref{eq:FP_tilde}} with the
time-dependent colored-noise boundary condition \prettyref{eq:Boundary_conditions_static},
which is explicitly shown in \citep{Schuecker14_arxiv}. Here we aim
for an even further reduction, recasting this time-dependent boundary
condition into a static condition by considering a shift in the threshold
\begin{equation}
\tilde{\theta}=\theta+k\frac{\alpha}{2}.\label{eq:shift_threshold}
\end{equation}
We perform a Taylor expansion of the effective density
\begin{equation}
\tilde{P}(\tilde{\theta},s)=\tilde{P}(\theta,s)+k\frac{\alpha}{2}\partial_{y}\tilde{P}(\theta,s)+O(k^{2})=O(k^{2})\label{eq:general_result}
\end{equation}
using \prettyref{eq:Boundary_conditions_static} and $\partial_{y}\tilde{P}(\theta,s)=-2\nu_{y}^{(0)}(s)+O(k)$
to show that the density vanishes to order $k^{2}$. As a result,
to first order in $k$ the dynamic boundary condition \eqref{eq:Boundary_conditions_static}
can be rewritten as a perfectly absorbing (white-noise) boundary at
shifted $\tilde{\theta}$ \eqref{eq:shift_threshold}. For the particular
problem of the stationary MFPT this was already found by \citet{Klosek98}
and \citet{Fourcaud02} as a corollary deduced from the steady state
density rather than as the result of a generic reduction of a colored-
to a white-noise problem. It remained unclear from these earlier works,
whether and how it is possible to apply the effective FPE to time-dependent
problems. \textcolor{black}{We show above that the effective FPE \eqref{eq:FP_tilde}
that directly follows from a perturbation expansion of the flux operator
can be re-summed to an effective operator acting on a one-dimensional
density. This, together with the effective boundary condition }\eqref{eq:Boundary_conditions_static}\textcolor{black}{{}
or \eqref{eq:shift_threshold}, is the crucial step to reduce the
time-dependent problem driven by colored noise to an effective time-dependent
white-noise problem, rendering the transient properties of the system
accessible to an analytical treatment.} 

After escape certain physical systems reset the dynamic variable $y$
to a value $R$, while leaving the noise variable $z$ unchanged.
Note that the noise density at reset therefore does not follow the
marginalized density shown in \prettyref{fig:Illustration-of-boundary}.
Instead, due to the fire and reset rule, it is biased to strictly
positive noise values and in the limit of weak noise its approximate
form was calculated for the PIF neuron \citep{Lindner04_0229011}.
In the presence of such a reset our calculation (see \prettyref{app:Reset})
yields the additional boundary condition $\tilde{P}(R+,s)-\tilde{P}(R-,s)=k\alpha\nu_{y}^{(0)}(s)$,
which to first order in $k$ is equivalent to a white-noise boundary
condition $\tilde{P}(\tilde{R}+,s)-\tilde{P}(\tilde{R}-,s)=O(k^{2})$
at shifted reset $\tilde{R}=R+k\frac{\alpha}{2}$. The extension to
the reset boundary condition is a further generalization of the theory
making it applicable to systems that physically exhibit a reset, such
as excitable biological membranes. Besides, it is a standard computational
procedure to re-insert the trajectories directly after escape at a
reset value in order to derive the stationary escape rate (\citep{Farkas27_236},
reviewed in \citep{haenggi90_251}).

\section{LIF-neuron\label{sec:LIF-neuron}}

We now apply the theory to the LIF model exposed to filtered synaptic
noise.\textcolor{black}{{} The} corresponding system of coupled differential
equations \citep{Fourcaud02} 
\begin{eqnarray}
\tau\dot{V} & = & -V+I+\mu\label{eq:IAF_diffusion}\\
\tau_{s}\dot{I} & = & -I+\sigma\sqrt{\tau}\,\xi\nonumber 
\end{eqnarray}
describes the evolution of the membrane potential $V$ and the synaptic
current $I$ driven by a Gaussian white noise with mean $\mu$ and
variance $\sigma^{2}$. The white noise typically represents synaptic
events that arrive at high rate but small individual amplitude as
a result of a diffusion approximation \citep{Ricciardi99}. Note that
the system \prettyref{eq:IAF_diffusion} can be obtained from \prettyref{eq:diffeq_general}
by introducing the coordinates 
\begin{equation}
y=\frac{V-\mu}{\sigma};\ z=\frac{k}{\sigma}I;\ s=t/\tau\label{eq:Trafo}
\end{equation}
and the choice of $f(y,s)=-y$ as a linear function. 

In the first subsection we consider the static case showing how our
central result simplifies the derivation of the earlier obtained stationary
firing rate. Hereby we introduce an operator notation exploiting the
analogy of the LIF neuron model to the quantum harmonic oscillator.
In the second subsection we combine this operator notation and our
general result \prettyref{eq:general_result} to study the effect
of synaptic filtering on the transfer of a time-dependent modulation
of the input to the neuron to the resulting time-dependent modulation
of its firing rate, characterized to linear order by the transfer
function. A derivation containing additional intermediate steps is
included in \citep{Schuecker14_arxiv}.

\subsection{Stationary firing rate\label{sub:Stationary-firing-rate}}

For the LIF neuron model the white-noise Fokker-Planck equation \eqref{eq:FP_tilde}
is

\begin{eqnarray}
\partial_{s}\rho(x,s) & = & -\partial_{x}\Phi(x,s)\equiv\mathcal{L}_{0}\,\rho(x,s)\nonumber \\
\Phi(x,s) & = & -(x+\partial_{x})\,\rho(x,s),\label{eq:FP_orig_transformed-1}
\end{eqnarray}
where we used $x=\sqrt{2}y$, equation \eqref{eq:Trafo}, and the
density \foreignlanguage{english}{$\rho(x,s)\equiv\frac{1}{\sqrt{2}}\tilde{P}(x/\sqrt{2},s)$}.
In order to transform the right hand side into a Hermitian form we
follow \citep{Risken96} and factor-off the square root of the stationary
solution, i.e. \foreignlanguage{english}{$u(x)=e^{-\frac{1}{4}x^{2}}$}.
We obtain 
\begin{equation}
\partial_{s}q(x,s)=-a^{\dagger}a\,q(x,s)\,,\label{eq:FP_QM}
\end{equation}
where $q(x,s)=u^{-1}(x)\rho(x,s)$ and we define the operators $a\equiv\frac{1}{2}x+\partial_{x}$,
$a^{\dagger}\equiv\frac{1}{2}x-\partial_{x}$, satisfying $[a,a^{\dagger}]=1$
and $a^{\dagger}a(a^{\dagger})^{n}q_{0}=n(a^{\dagger})^{n}q_{0}$,
where $q_{0}$ is the stationary solution of \eqref{eq:FP_QM} obeying
$aq_{0}=0$. So $a^{\dagger}$ is the ascending, $a$ the descending
operator and $(a^{\dagger})^{n}q_{0}$ is the $n$-th eigenstate of
the system \eqref{eq:FP_QM} as in the quantum harmonic oscillator.
The flux $\Phi(x)$ transforms to 
\[
-(x+\partial_{x})u(x)\circ=-u(x)(\frac{1}{2}x+\partial_{x})\circ=-u(x)a\circ
\]
and thus the stationary solution exhibiting a constant flux $\nu$
between reset and threshold obeys the ordinary inhomogeneous linear
differential equation 
\begin{equation}
-ua\,q_{0}(x)=\tau\nu\,H(x-x_{R})H(x_{\theta}-x).\label{eq:stationary_flux-1-1}
\end{equation}
The latter equation can be solved by standard methods given the white-noise
boundary conditions $0=uq_{0}(x_{\theta})=uq_{0}(x_{R+})-uq_{0}(x_{R-})$.
The normalization of the resulting stationary density then already
leads to the known result \citep{Siegert51,Ricciardi77} for the firing
rate 
\begin{equation}
(\tau\nu_{0})^{-1}=\int_{x_{R}}^{x_{\theta}}u^{-2}(x)F(x)\,dx,\label{eq:Siegert}
\end{equation}
with $F=\sqrt{\frac{\pi}{2}}(1+\erf(\frac{x}{\sqrt{2}}))$. In order
to obtain the correction of the firing rate due to colored noise we
only have to shift the boundaries in this expression according to
our general result \prettyref{eq:shift_threshold}, i.e.
\begin{equation}
\{\theta,R\}\rightarrow\{\theta,R\}+\sqrt{\tau_{s}/\tau}\frac{\alpha}{2},\label{eq:shifted_boundaries}
\end{equation}
which yields

\begin{equation}
(\tau\nu)^{-1}=\int_{x_{R}+\frac{\alpha k}{\sqrt{2}}}^{x_{\theta}+\frac{\alpha k}{\sqrt{2}}}u^{-2}(x)F(x)\,dx,\label{eq:Siegert_shift}
\end{equation}
identical to the earlier found solution \citep{Brunel01_2186,Fourcaud02}.
In contrast to these works, we here obtain the expression without
any explicit calculation, since the shift of the boundaries emerges
from the generic reduction from colored to white noise (\prettyref{sec:Reduction-from-colored}).

\begin{figure}[H]
\begin{centering}
\includegraphics{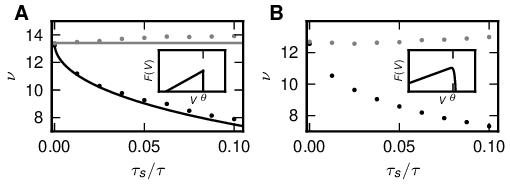}
\par\end{centering}

\caption{\textbf{\label{fig:Dependence-of-stationary}Dependence of stationary
firing rate of LIF model on synaptic filtering.} \textbf{A} Analytical
prediction \prettyref{eq:Siegert_shift} of stationary firing rate
$\nu$ (solid) in comparison to direct simulation (dots) for $V_{\theta}=20\protect\mV$,
$V_{r}=15\protect\mV$, $\mu=16.42\protect\mV$, $\sigma=4\protect\mV$,
$\tau=20\protect\ms$ (black) and shifted threshold and reset $\{V_{\theta},V_{R}\}\rightarrow\{V_{\theta},V_{R}\}-\sigma\sqrt{\tau_{s}/\tau}\frac{\alpha}{2}$
(gray). Enlargement of potential $F(V)=\frac{1}{2}V^{2}$ close to
perfectly absorbing boundary at $\theta$ shown in inset.\textbf{
B }Stationary firing rate $\nu$ for same parameters but with smooth
boundary shown in inset with $F(V)=\frac{1}{2}V^{2}-\Delta_{T}^{2}e^{(V-V_{\theta})/\Delta_{T}}$
and $\Delta_{T}=0.01$. Spike time $t_{\mathrm{s}}$ defined as $V(t_{\mathrm{s}})=30\protect\mV$.}
\end{figure}

The analytical expression \prettyref{eq:Siegert_shift} is in agreement
with direct simulations up to $k=\sqrt{0.1}$ (\prettyref{fig:Dependence-of-stationary}A).
All simulations were carried out with NEST \citep{nest_url}. The
firing rate shows a dependence on $\sqrt{\tau_{s}}$ , which becomes
obvious from a Taylor expansion of \prettyref{eq:Siegert_shift} in
$k$. The figure shows additional simulations in which the threshold
and the reset parameter in the neuron model are shifted according
to 
\begin{equation}
\{\theta,R\}\rightarrow\{\theta,R\}-\sqrt{\tau_{s}/\tau}\frac{\alpha}{2},\label{eq:inverse_shifted_boundaries}
\end{equation}
counteracting the shift \prettyref{eq:shift_threshold} caused by
the synaptic filtering. As predicted by the theory the firing rate
stays constant as the synaptic time constant is increased.

In conclusion the leading order correction to the stationary rate
or equivalently the MFPT stems only from the correct treatment of
the boundary conditions. In order to show that this effect is not
due to the simplified assumption of a perfectly absorbing boundary,
we next consider the case of a continuous boundary of the potential
in \prettyref{fig:Dependence-of-stationary}B. For concreteness we
choose the exponential integrate-and-fire neuron model. For a steep
falloff of the potential we still observe a $\sqrt{\tau_{s}}$ dependence
of the firing rate and more importantly are able to keep the firing
rate constant as we shift the location of the boundaries, indicating
that our central result holds true if the condition of a perfectly
absorbing threshold is relaxed. It is therefore adequate to approximate
the escape process by the system reaching a fixed threshold. Close
to this point, the density resembles the boundary layer solution for
the hard potential. For smoother potentials, the boundary layer will
successively be softened and its correction to order $\sqrt{\tau_{s}/\tau}$
diminishes \citep{Doering88}. This is in line with the recent observation
that the dominant correction to the transfer function of the exponential
integrate-and-fire model with a smooth falloff is of order $\tau_{s}/\tau$
\citep{alijani11_011919}.

\subsection{Transfer function\label{sub:Transfer-function}}

We now derive a novel first order correction for the input-output
transfer function up to moderate frequencies, which in the case of
the PIF model neuron has been shown to vanish \citep{Fourcaud02}.
This complements the earlier obtained limit for the modulation at
high frequencies \citep{Fourcaud02,Brunel01_2186}. 

We first simplify the derivation for white noise \citep{Brunel99,Lindner01_2934}
by exploiting the analogies to the quantum harmonic oscillator introduced
in \prettyref{sub:Stationary-firing-rate} and then study the effect
of colored noise. To linear order a periodic input with $\mu(t)=\mu+\epsilon\mu\,e^{i\omega t}$
and $\sigma^{2}(t)=\sigma^{2}+$\foreignlanguage{english}{$H\sigma^{2}e^{i\omega t}$}
modulates the firing rate $\nu_{0}(t)/v_{0}=1+n(\omega)e^{i\omega t}$,
proportional to the transfer function $n(\omega)$ to be determined.
With a perturbation ansatz for the modulated density \foreignlanguage{english}{$\rho(x,s)=\rho_{0}(x)+\rho_{1}(x,s)$}
and the separation of the time dependent part, $\rho_{1}(x)\,e^{i\omega\tau s}$
follows a second order ordinary linear differential equation $i\omega\tau\,\rho_{1}=\LN\rho_{1}+\LO\rho_{0}$.
Here $\LO=-G\,\partial_{x}+H\partial_{x}^{2}$ is the perturbation
operator, the first term of which originates from the periodic modulation
of the mean input with \foreignlanguage{english}{$G=\sqrt{2}\epsilon\mu/\sigma$,}
the second terms stems from the modulation of the variance. In operator
notation the perturbed FPE transforms to $(i\omega\tau+a^{\dagger}a)\,q_{1}=(G\,a^{\dagger}+H\,(a^{\dagger})^{2})\,q_{0}$
with the particular solution 
\[
q_{p}=\frac{G}{1+i\omega\tau}\,a^{\dagger}q_{0}+\frac{H}{2+i\omega\tau}\,(a^{\dagger})^{2}q_{0},
\]
where we used the properties of the ladder operators $a^{\dagger}a(a^{\dagger})q_{0}=(a^{\dagger})q_{0}$
and $a^{\dagger}a(a^{\dagger})^{2}q_{0}=2(a^{\dagger})^{2}q_{0}$.
We observe that the variation of $\mu$ contributes the first excited
state, the modulation of $\sigma^{2}$ the second. The homogeneous
solution $q_{h}$ can be expressed as a linear combination of parabolic
cylinder functions $U(i\omega\tau-\frac{1}{2},x)$, $V(i\omega\tau-\frac{1}{2},x)$
\citep{Abramowitz74,Lindner01_2934}. For white noise, the boundary
condition on $q_{1}=q_{h}+q_{p}$ is
\begin{eqnarray}
0 & = & q_{1}(x_{\theta})=q_{1}(x_{R+})-q_{1}(x_{R-})\label{eq:bc_q1}\\
 & \equiv & \left.q_{1}(x)\right|_{\{x_{R},x_{\theta}\}},\nonumber 
\end{eqnarray}
where we introduced a shorthand notation in the second line. The flux
due to the perturbation can be expressed with the transfer function
as a sum of two contributions, the first resulting from the unperturbed
flux operator acting on the perturbed density, the second from the
perturbed operator acting on the unperturbed density and consistently
neglecting the term of second order, i.e. 
\begin{equation}
\tau\nu_{0}\,n(\omega)=\left.u\left(-aq_{1}+(G+H\,a^{\dagger})q_{0}\right)\right|_{\{x_{R},x_{\theta}\}}.\label{eq:bc_flux}
\end{equation}
Knowing the particular solution yields four conditions, for the homogeneous
solution \foreignlanguage{english}{$\left.q_{h}\right|_{\{x_{R},x_{\theta}\}}$}
and its derivative \foreignlanguage{english}{$\left.\partial_{x}q_{h}\right|_{\{x_{R},x_{\theta}\}}$}
that determine the homogeneous solution on the whole domain as well
as the transfer function arising from the solvability condition as
\begin{equation}
n(\omega)=\underbrace{\frac{G}{1+i\omega\tau}\,\frac{\left.\Phi_{\omega}^{\prime}\right|_{x_{\theta}}^{x_{R}}}{\left.\Phi_{\omega}\right|_{x_{\theta}}^{x_{R}}}}_{\equiv n_{G}(\omega)}+\underbrace{\frac{H}{2+i\omega\tau}\,\frac{\left.\Phi_{\omega}^{\prime\prime}\right|_{x_{\theta}}^{x_{R}}}{\left.\Phi_{\omega}\right|_{x_{\theta}}^{x_{R}}}}_{\equiv n_{H}(\omega)},\label{eq:transfer_final}
\end{equation}
where $x_{\{R,\theta\}}=\sqrt{2}\frac{\{V_{R},V_{\theta}\}-\mu}{\sigma}$
and we introduced $\Phi_{\omega}(x)=u^{-1}(x)\,U(i\omega\tau-\frac{1}{2},x)$
as well as $\Phi_{\omega}^{\prime}=\partial_{x}\Phi_{\omega}$ to
obtain the known result \citep{Brunel99,Brunel01_2186,Lindner01_2934}.

\begin{figure}[H]
\centering{}\includegraphics{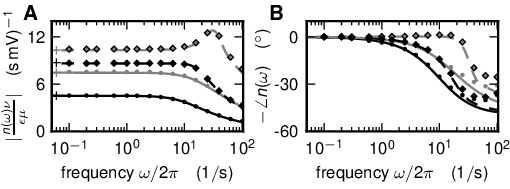}\caption{\textbf{\label{fig:working point}Colored-noise transfer function
of LIF model in different regimes.} Absolute value (\textbf{A}) and
phase (\textbf{B}) of the transfer function (vertical) for $V_{\theta}=20\protect\mV$,
$V_{r}=15\protect\mV$, $\tau_{m}=20\protect\ms$, $\tau_{s}=0.5\protect\ms$,
$\sigma=4\protect\mV$ (solid), $\sigma=1.5\protect\mV$ (dashed)
as a function of frequency (log-scaled horizontal axis). The mean
input $\mu$ was adapted to obtain different firing rates $\nu=10\protect\Hz$
(black) and $\nu=30\protect\Hz$ (gray). Analytical prediction $\tilde{n}_{G}$
(solid curves, \eqref{eq:transfer_final} with boundaries shifted
according to \eqref{eq:shifted_boundaries}), direct simulations (dots,
diamonds), and zero frequency limit \foreignlanguage{english}{$\frac{d\nu}{d\mu}$}
(crosses, obtained from \eqref{eq:Siegert_shift}).}
\end{figure}

\begin{figure}[H]
\centering{}\includegraphics{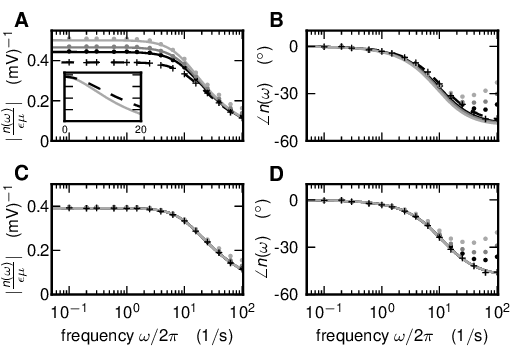}\caption{\textbf{\label{fig:Comparision-white-colored}Comparison between colored-
and white-noise transfer function.} Upper row: Absolute value (\textbf{A})
and phase shift (\textbf{B}) for $\sigma=4\protect\mV$, white noise
(dashed), colored noise $\tau_{s}\in[0.5,1,2]\protect\ms$ (from black
to gray) and $\tau_{s}=2\protect\ms$ normalized to zero frequency
limit of white noise (gray, inset). Lower row (\textbf{C},\textbf{D}):
same as (\textbf{A},\textbf{B}) but threshold and reset shifted $\{\theta,R\}\rightarrow\{\theta,R\}-\sqrt{\tau_{s}/\tau}\frac{\alpha}{2}$
to maintain constant firing rate\label{fig:sigma_comparision}. Display
and other parameters as in \prettyref{fig:working point}.}
\end{figure}

With the general theory developed above we directly obtain an approximation
for the colored-noise transfer function $n_{\mathrm{cn,G}}$, replacing
$x_{\{R,\theta\}}\rightarrow x_{\{\tilde{R},\tilde{\theta}\}}$ in
the white-noise solution \prettyref{eq:transfer_final}, denoted by
$\tilde{n}_{G}$. Here we only consider a modulation of the mean $\mu$,
because it dominates the response properties and briefly discuss a
modulation of the variance $\sigma^{2}$ in the next section. Note
that here the modulation enters the equation for $V$. If one is interested
in the linear response of the system with respect to a perturbation
of the input to $I$, as it appears in the neural context due to synaptic
input, one needs to take into account the additional low pass filtering
$\propto(1+i\omega\tau_{s})^{-1}$, which is trivial.

A Taylor expansion of the resulting function $\tilde{n}_{G}$ around
the original boundaries $x_{\{R,\theta\}}$ reveals the first order
correction in $k$
\begin{equation}
\begin{aligned}n_{\mathrm{cn,G}}(\omega)= & n_{G}(\omega)\\
 & +\sqrt{\frac{\tau_{s}}{\tau}}\frac{\alpha}{\sqrt{2}}\,\frac{G}{1+i\omega\tau}\left(\frac{\Phi_{\omega}^{\prime\prime}\vert_{x_{\theta}}^{x_{R}}}{\Phi_{\omega}\vert_{x_{\theta}}^{x_{R}}}-\left(\frac{\Phi_{\omega}^{\prime}\vert_{x_{\theta}}^{x_{R}}}{\Phi_{\omega}\vert_{x_{\theta}}^{x_{R}}}\right)^{2}\right),
\end{aligned}
\label{eq:Transfer_colored_lin}
\end{equation}
valid for arbitrary noise intensity $\sigma$ entering the expression
via the boundaries $x_{\{R,\theta\}}$. The first correction term
is similar to the $H$-term in \prettyref{eq:transfer_final} indicating
that colored noise has a similar effect on the transfer function as
a modulation of the variance \citep{Lindner01_2934} in the white-noise
case. In the high frequency limit this similarity was already found:
modulation of the variance leads to finite transmission at infinite
frequencies in the white-noise system \citep{Lindner01_2934}. The
same is true for modulation of the mean in the presence of filtered
noise \citep{Brunel01_2186}. However, for infinite frequencies our
analytical expression behaves differently. The two correction terms
in the second line of \prettyref{eq:Transfer_colored_lin} cancel
each other, since $\Phi_{\omega}^{\prime\prime}\rightarrow(i\omega\tau)^{2}\,\Phi_{\omega}$
and $\Phi_{\omega}^{\prime^{2}}\rightarrow(i\omega\tau)^{2}\,\Phi_{\omega}^{2}$.
Thus the transfer function decays to zero as in the white-noise case.
This deviation originates from neglecting the time derivative on the
left hand side of \prettyref{eq:FP_boundary_variable} that is of
order $k^{2}$, which is true only up to moderate frequencies $\omega\tau k\ll1$.
The high frequency limit $\omega\tau k\gg1$ is known explicitly \citep{Brunel01_2186},
but it was shown that the finite high-frequency transmission is due
to the artificial hard threshold in the LIF model \citep{Naundorf03,Fourcaud03_11640}.
\prettyref{fig:working point} shows a comparison of the analytical
prediction \prettyref{eq:Transfer_colored_lin} to direct simulations
for different noise levels $\sigma$ and firing rates controlled by
the mean $\mu$. The absolute value of the analytical result is in
agreement with simulations for the displayed range of frequencies.
Above $100\Hz$ deviations occur as expected. However, these are less
important, since our theory predicts the response properties well
in the frequency range where transmission is high. The deviations
are more pronounced and already observed at lower frequencies in the
phase shift. At high firing rate and low noise, the neuron is mean
driven and exhibits a resonance at its firing frequency, again well
predicted by the analytical result. The upper row in \prettyref{fig:Comparision-white-colored}
shows a comparison to the white-noise case. A qualitative change can
be observed: synaptic filtering on the one hand increases the dc-susceptibility
in contrast to the decreased firing rate (\prettyref{fig:Dependence-of-stationary}A).
For $\tau_{s}=2\ms$ this increase is already $25\%$, implying that
this correction is not negligible even for short time constants $\tau_{s}$.
On the other hand the cutoff frequency (inset in panel A) is reduced,
a behavior which is not expected from the existing knowledge that
colored noise enhances the transmission in the high frequency limit.
Our theory fully explains these qualitative changes by a shift in
the reset and the threshold analogously to the stationary case (\prettyref{fig:Dependence-of-stationary}).
This is shown explicitly in panels C and D: for different synaptic
time constants the response in the low frequency regime is not altered
in comparison to the white-noise case if reset and threshold are adapted
according to $\{\theta,R\}\rightarrow\{\theta,R\}-\sqrt{\tau_{s}/\tau}\frac{\alpha}{2}$
to compensate for the effect of colored noise.

\section{Balanced Random Network\label{sec:Balanced-Random-Network}}

\begin{figure}
\begin{centering}
\includegraphics{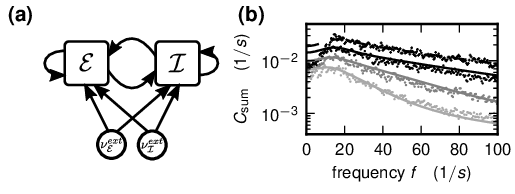}
\par\end{centering}

\caption{\label{fig:spectra}\textbf{Effect of synaptic filtering on power
spectra of the population activity in a balanced random network model.
A} Illustration of network architecture. \textbf{B} Theoretical prediction
\prettyref{eq:spectra_sum} (solid curves) vs. simulations (symbols)
for white noise (black dashes, crosses) and different values of $\tau_{s}\in[0.5,2,4]\protect\ms$
(from black to gray, dots). The network operates at the set point
$\mu=16.42\protect\mV,$ $\sigma=4\protect\mV$. Neuron parameters
are identical to the ones used in \prettyref{fig:sigma_comparision},
except a reset of $V_{r}=0$. Other network parameters: $N=10,000,\ \gamma=0.25,\ g=5,\ J=0.05\protect\mV,\ d=1.5\protect\ms$.
Power spectra obtained from a simulation of $T=100\,s$ and smoothed
by moving average (frame size $1\protect\Hz$). }
\end{figure}

On the network level, the theory of fluctuations (reviewed in \citep{Grytskyy13_131})
relies on the transfer function as well as the analysis of the stability
of networks \citep{Abbott93_1483} and of emerging oscillations \citep{Brunel99}.
With the results from the previous section these collective properties
of recurrent networks become analytically accessible for the biologically
relevant case of synaptic filtering. We here consider a network consisting
of $N_{\Ex}=N$ excitatory and $N_{\In}=\gamma N$ inhibitory LIF-model
neurons (\prettyref{sec:LIF-neuron}). The architecture is similar
to the one studied in \citep{Brunel00} with the extension to synaptic
filtering. Each neuron $i$ obeys the coupled set of differential
equations \foreignlanguage{english}{
\begin{eqnarray}
\tau\frac{dV_{i}}{dt} & = & -V_{i}+I_{i}(t)\nonumber \\
\tau_{s}\frac{dI_{i}}{dt} & = & -I_{i}+\tau\sum_{j=1}^{N}J_{ij}\,s_{j}(t-d)\label{eq:diffeq_iaf}
\end{eqnarray}
}corresponding to \prettyref{eq:IAF_diffusion}, but with input provided
by presynaptic spike trains $s_{i}(t)=\sum_{k}\delta(t-t_{k}^{i})$,
where the $t_{k}^{i}$ mark the time points at which neuron $i$ emits
an action potential, i.e. where $V_{i}$ exceeds the threshold and
$d$ denotes the synaptic delay. We use the same neuron parameters
as in \prettyref{fig:sigma_comparision}, except a reset of $V_{r}=0$
to reduce the impact of an incoming synaptic event, avoiding nonlinear
effects. The synaptic weight is $J_{ij}=J$ for excitatory and $J_{ij}=-gJ$
for inhibitory synapses. Each excitatory neuron receives $K_{\Ex}=pN_{\Ex}$
excitatory inputs, $p$ being the connection probability, and $K_{\In}=\gamma K_{\Ex}$
inhibitory inputs drawn randomly and uniformly from the respective
populations. Additionally, the neurons receive excitatory and inhibitory
external Poisson drive with rates $\nu_{\Ex,\In}^{\text{ext}}$ to
fix the set point defined by the mean $\mu=\tau K_{\Ex}J(1-\gamma g)\nu+\mu_{\mathrm{ext}}$
and the standard deviation $\sigma=\sqrt{\tau K_{\Ex}J^{2}(1+\gamma g^{2})\nu+\sigma_{\mathrm{ext}}^{2}}$
of the fluctuating input to a cell. Due to homogeneity the excitatory
and inhibitory neurons can be summarized into populations with activity
$\boldsymbol{s}=\left(\begin{array}{cc}
s_{\Ex}(t), & s_{\In}(t)\end{array}\right)^{T}$, where $s_{\alpha}(t)=\frac{1}{N_{\alpha}}\sum_{i\in\alpha}s_{i}(t),\ \alpha\in\{\Ex,\In\}$.
In this two-dimensional representation the connectivity can be expressed
by the in-degree matrix 
\[
\boldsymbol{K}=K\left(\begin{array}{cc}
1 & \gamma\\
1 & \gamma
\end{array}\right).
\]
A sketch of the network architecture is shown in \prettyref{fig:spectra}A.

The autospectra and cross-spectra $\boldsymbol{C}(\omega)$ of the
population activity can be determined by mapping the system to a linear
rate model \citep{Grytskyy13_131}. The solution is
\begin{equation}
\boldsymbol{C}(\omega)=\boldsymbol{P}(\omega)\,\boldsymbol{D}\,\boldsymbol{P}^{T}(-\omega)\label{eq:spectra_2D}
\end{equation}
with the diagonal matrix $\boldsymbol{D}_{\alpha\beta}=\delta_{\alpha\beta}\,\nu_{\alpha}\,N_{\alpha}^{-1}$
and the propagator matrix 
\[
P_{\alpha\beta}(\omega)=e^{-i\omega d}K_{\alpha\beta}H_{\alpha\beta}(\omega).
\]
Here $H_{\alpha\beta}$ is the weighted transfer function
\[
H_{\alpha\beta}(\omega)=\tau\nu_{\alpha}J_{\alpha\beta}\tilde{n}_{G}(\omega)\,\frac{1}{1+i\omega\tau_{s}}+\tau\nu_{\alpha}J_{\alpha\beta}^{2}\tilde{n}_{H}(\omega),
\]
with $\nu_{\alpha}$ given by \prettyref{eq:Siegert_shift} and identifying
$G=\sqrt{2}/\sigma$, $H=1/\sigma^{2}$ in $\tilde{n}_{G}$ and $\tilde{n}_{H}$
given by \eqref{eq:transfer_final}, where we use shifted boundaries
\eqref{eq:shifted_boundaries} to account for the colored noise. Here,
we also take into account the contribution to the colored-noise transfer
function $\tilde{n}_{H}(\omega)$ originating from a modulation of
the variance. Its form is again obtained from a shift of the boundaries
in the white noise solution $n_{H}$, since the general method of
reduction (\prettyref{sec:Reduction-from-colored}) holds true also
for a time-dependent variance $\sigma^{2}(t)=\sigma^{2}+H\sigma^{2}e^{i\omega t}$
as shown in \prettyref{app:modulation_sigma}. 

From \prettyref{eq:spectra_2D} we obtain the autospectra of the network
activity $s_{\text{sum}}=\frac{1}{N}\sum_{i\in\Ex,\In}s_{i}$ by a
weighted average over the matrix entries
\begin{equation}
C_{\text{sum}}=\frac{1}{N}(N_{\Ex}^{2}C_{\Ex\Ex}+N_{\Ex}N_{\In}C_{\Ex\In}+N_{\In}N_{\Ex}C_{\In\Ex}+N_{\In}^{2}C_{\In\In}).\label{eq:spectra_sum}
\end{equation}
The analytical prediction \prettyref{eq:spectra_sum} is in excellent
agreement with the spectra obtained in direct simulations of the network
(\prettyref{fig:spectra}). The shape of the power spectra is well
captured for different values of $\tau_{s}$ and the theory predicts
the suppression of fluctuations at frequencies in the low ($30$-$60\Hz$)
and the high gamma range ($60-200\Hz$) caused by synaptic filtering.
These frequency bands are related to task dependent activity studied
in animals \citep{Ray08_1529} and humans \citep{Ball08_302}. Although
the synaptic time constants are small compared to the membrane time
constant ($k\le0.44$), the influence on the network dynamics is strong.
We here present for the first time an analytical argument explaining
the effect. The observed deviations at low frequencies are expected:
The theory in \citep{Grytskyy13_131} assumes the autocorrelation
of single neurons to be $\delta$-shaped. This is an approximation
likely to be violated, since after reset the membrane potential needs
time to recharge, resulting in a dip of the autocorrelation around
zero corresponding to a reduction of power at low frequencies. However,
the spectrum of the population activity is dominated by cross-correlations
resulting in a good overall agreement between theory and simulation.
\citet{dummer14} show that the influence of the cross-correlations
on the single neuron's auto-correlation is negligible and developed
an iterative procedure which numerically solves for the self-consistent
solution of the auto-correlation. One could incorporate this approach
into the theory presented here, replacing the simplified assumption
of a $\delta$-shaped auto-correlation with its self-consistent solution.

\section{Discussion}

In this manuscript we investigate the impact of realistic colored
noise on the escape of a Brownian particle from a meta-stable state.
Independent of the physical system, specified by the potential, the
leading order correction to the escape time solely comes about by
the effect of the colored noise in conjunction with the absorbing
boundary; there is no correction at the leading order to the Fokker-Planck
equation describing the bulk of the density away from the boundary.
This is reflected in our generic main result: A time-dependent system
driven by colored-noise can to first order be equivalently described
by a white-noise system with shifted boundary locations. The modification
of the boundary condition can be understood on physical grounds. An
absorbing boundary in a white-noise driven system results in a vanishing
density at the boundary, because the infinitely fast noise causes
any state sufficiently close to the boundary to immediately cross
this threshold. The situation is different if the noise has a band-limited
spectrum. The diffusive motion induced by the noise is slower, reducing
the rate of escape for states close to the boundary. Since the density
follows a continuity equation (it obeys conservation of probability),
its stationary magnitude results from an equilibrium between inflow
and outflow at any instant. A reduced outflow over the threshold is
therefore accompanied by elevated density at threshold, which is exactly
what is formally achieved by displacing the perfectly absorbing white-noise
boundary beyond the point of the physical threshold. This result is
in line with the recently observed phenomenon of noise-enhanced stability
of a meta-stable state that exhibits a shift of the critical initial
position as a function of the color of the noise \citep{Fiascanaro09},
similar to the shift of the absorbing boundary. 

As discussed in \prettyref{sub:Transfer-function} the general reduction
presented here is valid up to moderate frequencies, while for high
frequencies the time derivative on the left hand side of \prettyref{eq:FP_boundary_variable}
cannot be treated as a second order perturbation in $k$. The presented
theory could be extended by absorbing the time derivative into the
operator $\tilde{L}=L-i\omega\tau$. However, this significantly complicates
the form of the outer as well as the boundary-layer solution and the
matching between the two. Additionally, the theory could be generalized
to multiplicative noise, i.e. an arbitrary factor $g(y)$ in front
of the noise $\xi$ in \eqref{eq:diffeq_general}. 

In \prettyref{sec:LIF-neuron} we apply the central result to the
LIF neuron model and obtain an approximation for its transfer function
in the presence of colored noise. We show that for biologically relevant
parameters the theory presents a viable approximation. In deriving
the underlying white-noise transfer function we exploit the analogy
between diffusion processes and quantum mechanics by transforming
the Fokker-Planck equation of the LIF model to the Hamiltonian of
the quantum harmonic oscillator. Besides the technical advantage of
using the established operator algebra, this formal analogy exposes
that the two components of the transfer function, due to a modulation
of the mean input and due to a modulation of the incoming fluctuations,
are related to the first and second excited state, respectively, of
the quantum harmonic oscillator. These two contributions to the neuronal
signal transmission have been studied theoretically \citep{Lindner01_2934}
and experimentally \citep{Silberberg04_704} and the exposed formal
analogy casts a new light on this known result. The analytical expression
\prettyref{eq:Transfer_colored_lin} of the transfer function for
the colored noise case has immediate applications: The transmission
of correlated activity by pairs of neurons exposed to common input
\citep{Shea-Brown08,DeLaRocha07_802}, can now be studied in a time-resolved
manner. Further it enables us to study the effect of synaptic filtering
on the oscillatory properties of recurrent networks, as shown in \prettyref{sec:Balanced-Random-Network}.
More generally, the result allows researchers to map out the phase
space (including regions of stability, and the emergence of oscillations)
of recurrent neuronal networks, analogous to the white-noise case
\citep{Brunel00}, also in presence of biologically realistic synaptic
filtering. This is in particular important as contemporary network
models in neuroscience tend to become more and more complex, featuring
multiple neuronal populations on the meso- and the macroscopic scale
\citep{Potjans14_785,Schmidt14}. Having an analytical description
at hand allows the identification of sub-circuits critical for the
emergence of oscillatory activity and fosters the systematic construction
of network models that are in line with experimental observations.
In connection with the recently proposed method of linear stability
analysis of pulse-coupled oscillators on the basis of perturbations
of the inter-event density \citep{Farkhooi15_038103}, our modified
boundary condition can replace the white noise boundary condition
\citep[their eq. (9)]{Farkhooi15_038103}, to obtain an approximation
for the case of colored noise avoiding the need to extend their method
to a two-dimensional Fokker-Planck equation.

Excitable systems are ubiquitous in nature and a large body of literature
discusses their properties. In particular the escape from a meta-stable
state activated by colored noise appears in diverse areas of the natural
sciences, including semiconductor physics, physical chemistry, genetics,
hydrodynamics and laser physics \citep{Moss89_2,Moss89_3}. The simplicity
of the presented method, reducing this general problem to the well
understood escape from a potential driven by white-noise, but with
displaced boundaries, opens a wide field of applications, where the
effect of realistic colored noise can now be accessed in a straight-forward
manner.
\begin{acknowledgments}
The authors thank Hannah Bos for substantially contributing to the
theory and the simulation code used in \prettyref{sec:Balanced-Random-Network}.
This work was partially supported by Helmholtz young investigator's
group VH-NG-1028, Helmholtz portfolio theme SMHB, J\"ulich Aachen
Research Alliance (JARA), EU Grant 269921 (BrainScaleS), and EU Grant
604102 (Human Brain Project, HBP). We thank two anonymous referees
for their constructive comments which helped us to substantially improve
the manuscript. We thank Benjamin Lindner and Nicolas Brunel for helpful
comments on an earlier version of the manuscript.
\end{acknowledgments}

\appendix

\section{Boundary condition at reset\label{app:Reset}}

Certain physical systems, such as biological membranes, exhibit a
reset to a smaller value by assigning $y\leftarrow R$ after $y$
has escaped over the threshold. This corresponds to the flux escaping
at threshold $\theta$ being re-inserted at reset $R$. The corresponding
boundary condition is
\begin{eqnarray*}
\nu_{y}(\theta,z,s) & = & \frac{z}{k}Q(\theta,z,s)\\
 & = & S_{y}\left(Q(R+,z,s)-Q(R-,z,s)\right).
\end{eqnarray*}
With the half-range boundary condition $\frac{z}{k}\,Q(\theta,z,s)=0\ \forall z<0$
we get 
\begin{eqnarray*}
\left(f(R,z,s)-f(\theta,z,s)+\frac{z}{k}\right)\left(Q(R+,z,s)-Q(R-,z,s)\right)\\
=0\quad\forall z<0,
\end{eqnarray*}
from which we conclude
\begin{equation}
0=Q(R+,z,s)-Q(R-,z,s)\quad\forall z<0.\label{eq:Q_reset_boundary_layer}
\end{equation}
We now consider the boundary layer at reset described with the rescaled
coordinate $r=\frac{y-R}{k}$. We define the boundary layer solution
\begin{eqnarray*}
Q^{B}(r,z,s) & = & Q^{+}(y(r),z,s)-Q^{-}(y(r),z,s)\\
 & \equiv & \Delta Q(y(r),z,s),
\end{eqnarray*}
where we introduced two auxiliary functions $Q^{+}$ and $Q^{-}$:
here $Q^{+}$ is a continuous solution of \prettyref{eq:FP_Q_general}
on the whole domain and, above reset, agrees to the solution that
obeys the boundary condition at reset. Correspondingly, the continuous
solution $Q^{-}$ agrees to the searched-for solution below reset.
With \prettyref{eq:Q_reset_boundary_layer} it follows $Q^{B}(0,z,s)=0\quad\forall z<0$.
Thus we have the same boundary condition as in the boundary layer
at threshold \prettyref{eq:BC_boundary_layer}. Therefore the calculations
\prettyref{eq:order1_boundary}-\prettyref{eq:Boundary_conditions_static}
can be performed analogously, whereby on has to use the continuity
of the white-noise solution at reset, i.e. $\Delta Q^{(0)}(R,s)=0$.
The result is 
\[
\Delta Q_{0}^{(1)}(R,s)=\tilde{P}(R+,s)-\tilde{P}(R-,s)=\alpha\nu_{y}^{(0)}(s),
\]
a time-dependent boundary condition for the jump of the first-order
correction of the outer solution at reset proportional to the probability
flux $\nu_{y}^{(0)}$ of the white-noise system.

\section{Modulation of the noise amplitude\label{app:modulation_sigma}}

We here show that the reduction from the colored-noise system to the
white-noise system presented in \prettyref{sec:Reduction-from-colored}
can be performed analogously in the case of a temporally modulated
noise. To this end we start with the system of equations

\begin{eqnarray*}
\frac{dy}{ds} & = & f(y,s)+\frac{z}{k}\\
k\frac{dz}{ds} & = & -\frac{z}{k}+\eta(s)\xi,
\end{eqnarray*}
where $\eta(s)$ is a time-dependent positive ($\eta(s)>0\quad\forall s$)
prefactor modulating the amplitude of the noise. Rescaling both coordinates
with the amplitude of the noise $\tilde{y}(s)=y(s)/\eta(s)$ and $\tilde{z}(s)=z(s)/\eta(s)$,
the temporal derivatives transform to $\frac{dy}{ds}=\frac{d}{ds}(\tilde{y}\eta)=\eta\frac{d\tilde{y}}{ds}+\tilde{y}\frac{d\eta}{ds}$
and $\frac{dz}{ds}=\frac{d}{ds}(\tilde{z}\eta)=\eta\frac{d\tilde{z}}{ds}+\tilde{z}\frac{d\eta}{ds}$.
After division by $\eta(s)$ and using $\frac{1}{\eta}\frac{d\eta}{ds}=\frac{d\ln\eta}{ds}$,
we get
\begin{eqnarray*}
\frac{d\tilde{y}}{ds} & = & \underbrace{f(\eta\tilde{y},s)-\frac{d\ln\eta}{ds}\tilde{y}}_{\equiv\tilde{f}(\tilde{y},s)}+\frac{\tilde{z}}{k}\\
k\frac{d\tilde{z}}{ds} & = & -\frac{\tilde{z}}{k}-k\frac{d\ln\eta}{ds}\,\tilde{z}+\xi,
\end{eqnarray*}
where we defined the function $\tilde{f}$ appearing in the modified
drift term for $\tilde{y}$. The corresponding Fokker-Planck equation
(cf. \eqref{eq:FP_2D} for case with noise of constant amplitude)
is
\begin{align}
k^{2}\partial_{s}P & =\partial_{\tilde{z}}\left(\frac{1}{2}\partial_{\tilde{z}}+\tilde{z}+k^{2}\,\tilde{z}\,\frac{d\ln\eta}{ds}\right)\,P-k^{2}\partial_{\tilde{y}}\tilde{S}_{\tilde{y}}\,P\nonumber \\
\tilde{S}_{\tilde{y}} & =\tilde{f}(\tilde{y},s)+\frac{\tilde{z}}{k}.\label{eq:FP_general_time_noise}
\end{align}
By comparing to \eqref{eq:FP_2D} we observe that the only formal
difference is the additional term $k^{2}\,\tilde{z}\,\frac{d\ln\eta}{ds}$.
Factoring-off the stationary solution of the fast part $P=Q\,\frac{e^{-\tilde{z}^{2}}}{\sqrt{\pi}}$
with
\begin{eqnarray*}
 &  & \partial_{\tilde{z}}\left(\tilde{z}\frac{d\ln\eta}{ds}\frac{e^{-\tilde{z}^{2}}}{\sqrt{\pi}}Q\right)\\
 & = & \frac{e^{-\tilde{z}^{2}}}{\sqrt{\pi}}\,\frac{d\ln\eta}{ds}\left(1-2\tilde{z}^{2}+\tilde{z}\partial_{\tilde{z}}\right)Q
\end{eqnarray*}
we obtain (cf. \eqref{eq:FP_Q_general})

\begin{eqnarray}
k^{2}\partial_{s}Q & = & LQ-k\tilde{z}\partial_{\tilde{y}}Q\label{eq:FP_Q_time_noise}\\
 &  & -k^{2}\left(\partial_{\tilde{y}}\,\tilde{f}(\tilde{y},s)-\frac{d\ln\eta}{ds}\left(1-2\tilde{z}^{2}+\tilde{z}\partial_{\tilde{z}}\right)\right)Q.\nonumber 
\end{eqnarray}
The additional term affects the perturbation expansion \eqref{eq:perturb_expansion_general}
at second order. Since $Q^{(0)}$ is independent of $\tilde{z}$ we
have $\partial_{\tilde{z}}Q^{(0)}=0$ which leads to
\begin{eqnarray*}
LQ^{(2)} & = & \partial_{s}Q^{(0)}+\tilde{z}\partial_{y}Q^{(1)}\\
 &  & +\left(\partial_{\tilde{y}}\,\tilde{f}(\tilde{y},s)-\frac{d\ln\eta}{ds}\left(1-2\tilde{z}^{2}\right)\right)Q^{(0)}.
\end{eqnarray*}
Due to $L\tilde{z}^{2}=1-2\tilde{z}^{2}$, the additional term in
the particular solution of the second order is $-\frac{d\ln\eta}{ds}\tilde{z}^{2}$.
This term does not contribute to the first order correction of the
marginalized probability flux in $\tilde{y}$-direction $\nu_{\tilde{y}}(\tilde{y},s)\equiv\int d\tilde{z}\,\frac{e^{-\tilde{z}^{2}}}{\sqrt{\pi}}S_{\tilde{y}}Q(\tilde{y},\tilde{z},s)=\sum_{n=0}^{1}k^{n}\,\nu_{\tilde{y}}^{(n)}(\tilde{y},s)+O(k^{2})$,
because the factor $\frac{\tilde{z}}{k}$ in $\tilde{S}_{\tilde{y}}$
causes a point-symmetric function in $\tilde{z}$ that vanishes after
integration over $\tilde{z}$ while the factor $\tilde{f}(\tilde{y})$
yields a contribution that is second order in $k$. Therefore the
effective flux has the same form as before \eqref{eq:effective_flux}
meaning that the one-dimensional Fokker-Planck equation \eqref{eq:FP_tilde}
with $f$ replaced with $\tilde{f}(\tilde{y},s)$ 
\[
\partial_{s}\tilde{P}=\partial_{\tilde{y}}\left(-\tilde{f}(\tilde{y},s)+\frac{1}{2}\partial_{\tilde{y}}\right)\,\tilde{P}
\]
holds for the case of time-modulated noise. The corresponding SDE
is

\[
\frac{d\tilde{y}}{ds}=\tilde{f}(\tilde{y},s)+\xi.
\]
Transforming back to the original coordinate $y=\eta\tilde{y}$ we
get
\[
\frac{dy}{ds}=f(y,s)+\eta(s)\xi,
\]
so that we obtain a SDE with time modulated white noise $\eta(s)\xi$.

To obtain the boundary condition, we follow the same calculation as
in \prettyref{sec:Reduction-from-colored} and transform the FPE \eqref{eq:FP_Q_time_noise}
to the shifted and scaled coordinates $\tilde{r}=\frac{\tilde{y}-\theta}{k}$,
and $\tilde{z}+kf(\theta,s)\to\tilde{z}$ which yields

\begin{equation}
\begin{aligned}k^{2}\partial_{s}Q^{B} & =LQ^{B}-\tilde{z}\partial_{\tilde{r}}Q^{B}\\
 & +G(\theta,\tilde{r},s,\tilde{z})\,k\,Q^{B}\\
 & +k^{2}\frac{d\ln\eta}{ds}\left(1-2\tilde{z}^{2}+\tilde{z}\partial_{\tilde{z}}\right)Q^{B}\\
 & +O(k^{3}),
\end{aligned}
\label{eq:FP_boundary_variable-2}
\end{equation}
with $Q^{B}(\tilde{r},\tilde{z},s)\equiv Q(\tilde{y}(\tilde{r}),\tilde{z},s)$
and the operator $G(\theta,\tilde{r},s,\tilde{z})=f(\theta,s)\partial_{\tilde{z}}-\partial_{\tilde{r}}\,(f(k\tilde{r}+\theta,s)-f(\theta,s))$.
Since the additional term $k^{2}\frac{d\ln\eta}{ds}\left(1-2\tilde{z}^{2}+\tilde{z}\partial_{\tilde{z}}\right)$
is of second order in $k$ it only affects the perturbative solution
of the second order in \eqref{eq:order1_boundary}, so the effective
boundary conditions \eqref{eq:Boundary_conditions_static} and \eqref{eq:shift_threshold}
also hold in the case of time-dependent noise. The reduction of the
colored-noise to the corresponding white-noise problem can therefore
be performed in the same way as for constant noise.
\end{document}